\documentclass[twocolumn,amsmath,amssymb,PRL]{revtex4}
\usepackage{color}
\usepackage{graphicx}

\input epsf
\begin{document}
\title {Anisotropic superfluidity in a dipolar Bose gas}
\author{Christopher Ticknor$^1$}
\author{Ryan M. Wilson$^2$}
\author{John L. Bohn$^2$} 
\affiliation{$^1$Theoretical Division, Los Alamos National Laboratory, Los Alamos, New Mexico 87545, USA}
\affiliation{$^2$JILA, NIST and Department of Physics, University of Colorado, Boulder, Colorado 80309, USA}
\date{\today}
\begin{abstract}
We study the superfluid character of a dipolar Bose-Einstein condensate (DBEC) in a quasi-two dimensional (q2D) geometry.  In particular, we allow for the dipole polarization to have some non-zero projection into the plane of the condensate so that the effective interaction is anisotropic in this plane, yielding an anisotropic dispersion for propagation of quasiparticles.  By performing direct numerical simulations of a probe moving through the DBEC, we observe the sudden onset of drag or creation of vortex-antivortex pairs at critical velocities that depend strongly on the direction of the probe's motion.  This anisotropy emerges because of the anisotropic manifestation of a roton-like mode in the system. 
\end{abstract}
\maketitle

A quintessential feature of a superfluid is its ability to support dissipationless flow, for example, when an object moves through a superfluid and experiences no drag force.  This, however, only occurs when the object is moving below a certain critical velocity; when it exceeds this critical velocity it dissipates energy into excitations of the superfluid, resulting in a net drag force on the object and the breakdown of superfluid flow. 

In many superfluids, such as dilute Bose-Einstein condensates (BECs) of atoms, this critical velocity is simply the speed of sound in the system, which is set by the density and the $s$-wave scattering length of the atoms.  However, for a dense superfluid such as liquid $^4$He, this is not the case.  In $^4$He, the critical velocity is set by a roton mode, corresponding to a peak in the static structure factor of the system at some finite, non-zero momentum, with a characteristic velocity that is considerably less than the speed of sound in the liquid.  This feature has been verified experimentally via measurements of ion-drift velocity in the fluid \cite{Allum}, thereby providing insight into the detailed structure of the system.

Interestingly, a BEC of dipolar constituents (DBEC) is also expected to possess a roton-like dispersion, in spite of existing in a dilute gaseous state~\cite{Santos03}.  Unlike the dispersion of $^4$He, the dispersion of a DBEC is highly tunable as a function of the condensate density and dipole-dipole interaction (ddi) strength.  Additionally, the DBEC is set apart from liquid $^4$He in that its interactions depend on how the dipoles are oriented in space.  Thus, the DBEC provides an ideal system to study the effects that anisotropies have on the bulk properties of a superfluid, such as its critical velocity.   Anisotropic dispersions have been predicted for a 1D lattice system of q2D DBECs~\cite{Wang}, periodically dressed BECs~\cite{Higbie} and for dipolar gases in a 2D lattice \cite{lattice}.  Additionally, anisotropic solitons have been predicted for dipolar gases \cite{soliton}.

In this Letter we consider a DBEC in a quasi-two-dimensional (q2D) geometry and allow for the dipoles to be polarized at a nonzero angle into this plane so that the in-plane interaction is anisotropic.  We perform numerical simulations of a probe moving through the DBEC. This probe experiences a sudden onset of drag at a certain velocity, the critical velocity $v_c$, which is very different depending on the direction of the probe's motion.  For perturbative probes the drag originates from the production of quasiparticles and $v_c$ is nearly equal to $v_L$, the critical velocity predicted by the dispersion of the DBEC.  In contrast, for a larger probe vortex-antivortex pairs are formed, usually leading to a critical velocity smaller than $v_L$ \cite{Frisch,win}.  For both cases we find that the critical velocity is larger in the direction of the dipole tilt than in the perpendicular direction. Interestingly, while the roton displays an anisotropic character, the speed of sound in the system remains isotropic.  Thus, we characterize the DBEC as an anisotropic superfluid while illuminating the crucial role that the roton plays in generating this anisotropic behavior. 
Importantly, such a system offers a path to study a bosonic system which has anisotropic collective behavior which is typically found in composite fermion systems, such as high $T_c$ $d$-wave Superconductors. 

We work in the q2D geometry by assuming that there is a strong one-dimensional (1D) harmonic trap in the $z$-direction, $U_\mathrm{1D}(z) = \frac{1}{2}m \omega_z^2 z^2$,  where $m$ is the bosonic mass and $\omega_z$ is the trapping frequency. This allows the condensate wavefunction to be written in the separable form $\Psi(\mathbf{r},t) = \chi(z)\psi(\boldsymbol\rho,t)$ where $\chi(z)$ is the ground state harmonic oscillator wavefunction.  By inserting this ansatz into the Gross-Pitaevskii equation (GPE), describing the dilute, zero-temperature DBEC, and integrating out the $z$-dependence, we derive the modified, time-dependent GPE for the q2D system,
\begin{equation}
\label{q2DGPE}
i\hbar \partial_t \psi = \left\{-{\hbar^2\over2m}\nabla^2 + V_p +g|\psi|^2 +g_d\Phi \right\}\psi ,
\end{equation}
where $\psi = \psi(\boldsymbol{\rho},t)$ is the in-plane condensate wavefunction, $V_p=V_p(\boldsymbol{\rho},t)$ is a time-dependent probe potential, $g=2\sqrt{2\pi} \hbar^2a_s/m l_z$ is the mean-field coupling for contact interactions, $a_s$ is the scattering length, $l_z=\sqrt{\hbar/m \omega_z }$ is the axial harmonic oscillator length, $g_d=d^2/\sqrt{2\pi} l_z$ is the ddi coupling and $g_d \Phi = g_d\Phi(\boldsymbol{\rho},t)$ is the mean-field potential due to the ddi, where $\Phi\left(\boldsymbol{\rho},t\right)=\frac{4\pi}{3}\mathcal{F}^{-1}[\tilde{n}(\mathbf{k},t) F(\frac{\mathbf{k} l_z}{\sqrt{2}})]$, by the convolution theorem. Here, $\mathcal{F}$ is the 2D Fourier transform operator and $\tilde{n}(\mathbf{k},t) = \mathcal{F}[n(\boldsymbol{\rho},t)]$.

The function $\frac{4\pi}{3} g_d F(\mathbf{q})$, where $\mathbf{q}\equiv \mathbf{k}l_z/\sqrt{2}$, is the $k$-space ddi for the q2D geometry. It has two contributions coming from polarization perpendicular or parallel to the direction of the dipole tilt, $F(\mathbf q)=\cos^2(\alpha)F_{\perp}(\mathbf q)+\sin^2(\alpha)F_{\parallel}(\mathbf q)$ where $\alpha$ is the angle between $\hat{z}$ and the polarization vector $\hat d$.  These contributions are $F_{\parallel}(\mathbf q)=-1+3\sqrt{\pi}(q_d^2/q)e^{q^2}\mbox{erfc}(q)$, where $\mathbf q_d$ is the wave vector along the direction of the projection of $\hat{d}$ onto the $x$-$y$ plane, erfc is the complementary error function and $F_{\perp}(\mathbf q)=2-3\sqrt{\pi}q e^{q^2}\mbox{erfc}(q)$.  

To simplify this problem, we rescale energies in units of the chemical potential, given by  $\mu^*={g n_0}\{1+\beta {4\pi\over3} (3\cos(\alpha)^2-1)\}$ for the unperturbed system~\cite{nath}.  Here, $\beta = g_d / g$.  This leads to characteristic units of length given by the coherence length $\xi^*=\hbar/\sqrt{m\mu^*}$; time $\tau^* =\hbar /\mu^*$; and velocity $c^*=\sqrt{\mu^*/m}$.  Additionally, we rescale the wavefunction $\psi\rightarrow\psi/\sqrt{n_0}$ where $n_0$ is the integrated 2D density of the unperturbed system.  The rescaled coupling constants are then $g^*=gn_0/\mu^*$ and $g_d^*=g_dn_0/\mu^*=\beta g^*$.  This formalism describes the q2D DBEC, including the ground (condensed) state and the dispersion relation that describes the system's quasiparticle spectrum.

The dispersion relation of a homogeneous q2D DBEC is given in Bogoliubov theory by~\cite{uwe}
\begin{equation}
\omega(\mathbf{k}) = \sqrt{\frac{k^4}{4}+k^2 g^*\left( 1+ \frac{4\pi}{3}\beta F\left( \frac{\mathbf{k}l_z}{\sqrt{2}} \right)\right)}.
\end{equation}
For $\alpha=0$ (polarization along the trap axis) this dispersion does not depend on the direction of the quasiparticle propagation.  However, for $\alpha\neq 0$, or for nonzero projection of $\hat{d}$ onto the $x$-$y$ plane, the direction of $\mathbf{k}$ becomes important in describing the quasiparticles of the system.  Landau famously used the dispersion relation of a Bose fluid to determine its superfluid critical velocity, defining the Landau critical velocity, $v_L= \min{[\omega(\mathbf{k})/k ]}$ \cite{Landau}.  As $v_L$ is given in terms of the dispersion relation $\omega(\mathbf{k})$, it then also depends on the direction of $\mathbf{k}$, and thus is an anisotropic quantity when $|\cos{(\alpha)}|<1$.

\begin{figure}
\vspace{-8pt}
\includegraphics[width=.9\columnwidth]{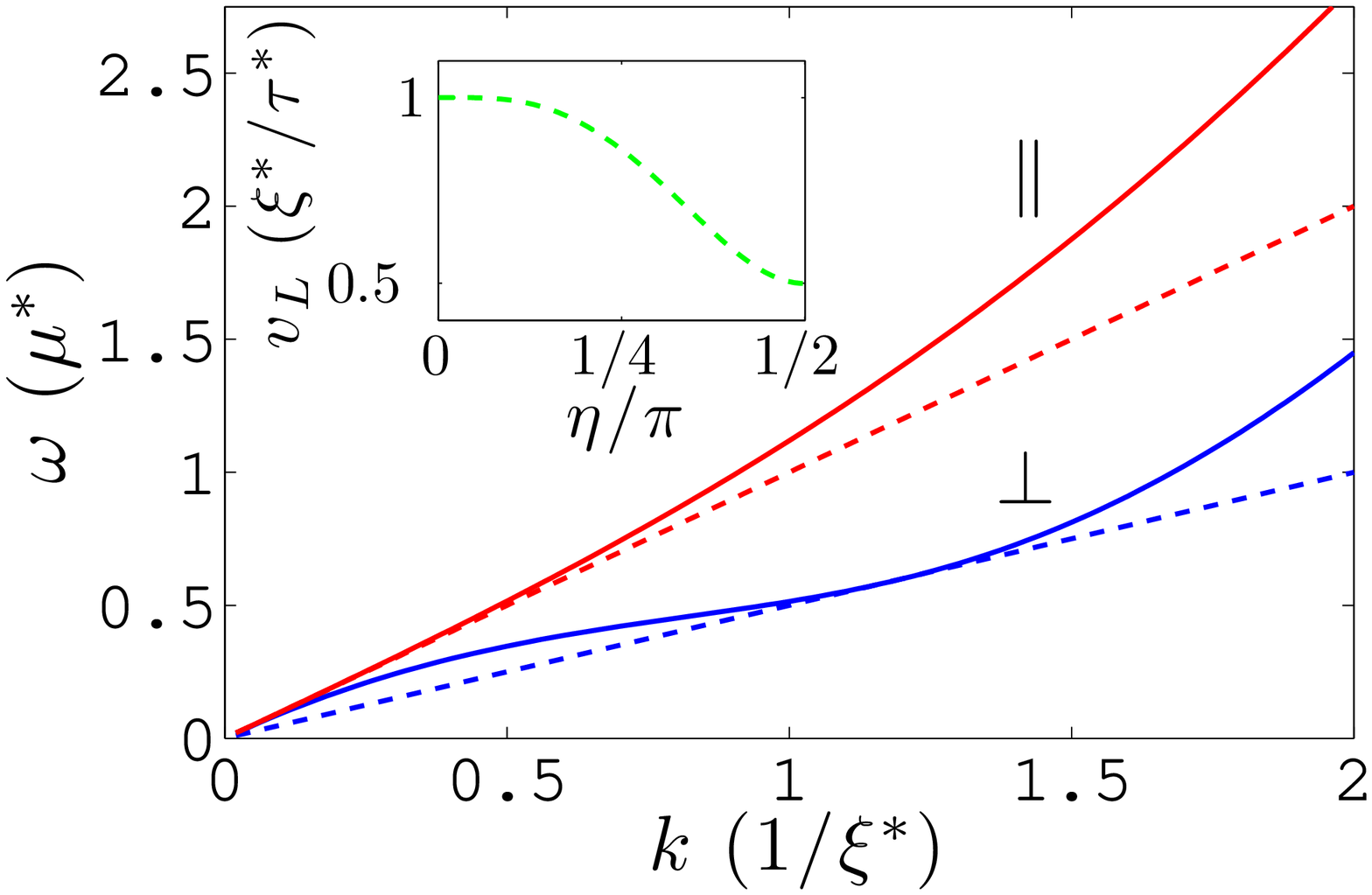} \\
\vspace{-12pt}
\includegraphics[width=0.95\columnwidth]{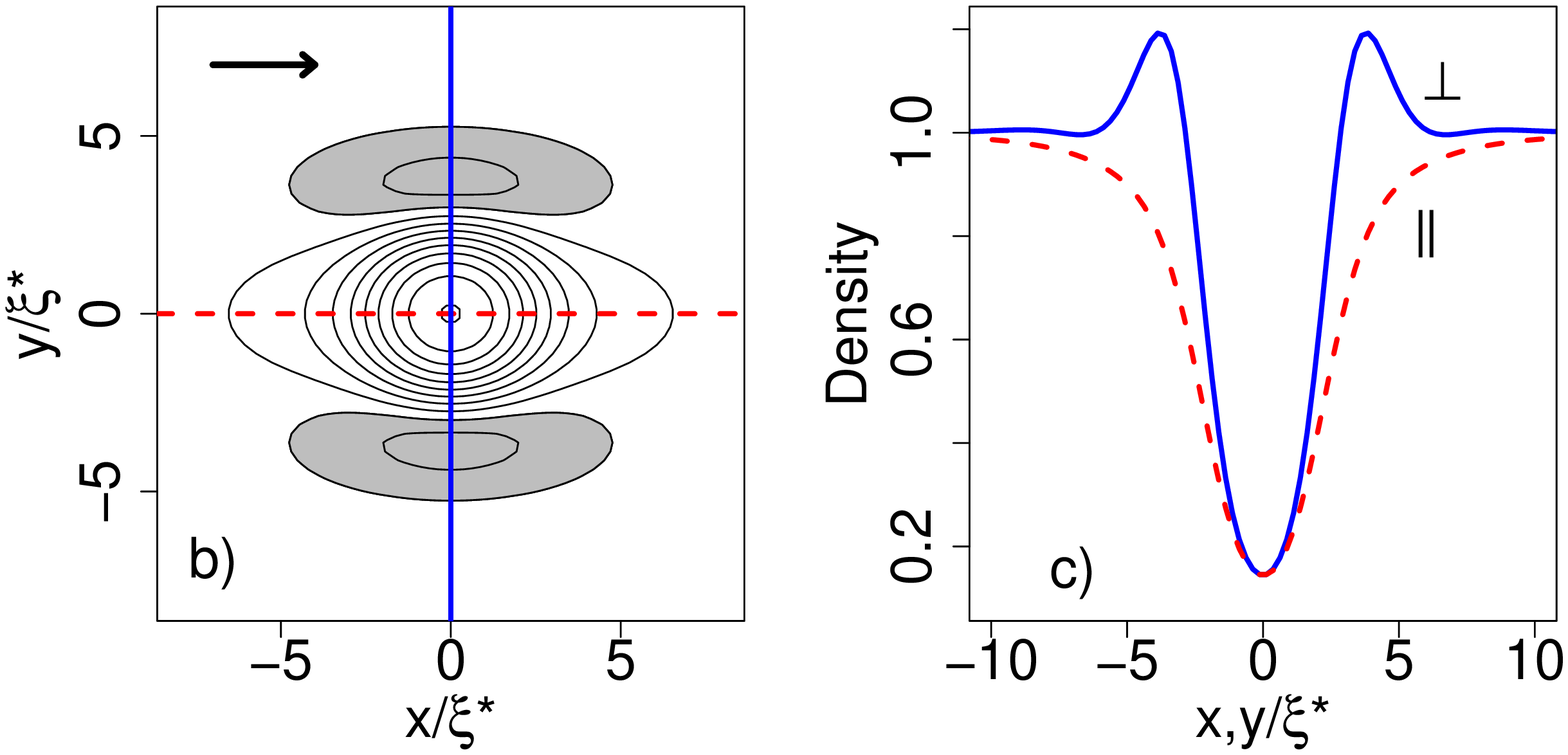}
\vspace{-4pt}
\caption{(color online)  (a) The Bogoliubov dispersions for $g^*=0.25$, $g_d^*=0.36$, $l_z/\xi^*=0.5$ and $\alpha = \pi/4$ for propagation perpendicular ($\perp$) to and parallel ($\parallel$) to the tilt of the dipoles, shown by the blue and red lines, respectively.  The dashed lines have slopes that are the Landau critical velocities ($v_L$) of the dispersions, while the inset shows $v_L$ as a smooth function of the angle $\eta$ between the $\parallel$ ($\eta/\pi = 0$) and $\perp$ ($\eta/\pi = 1/2$) propagation directions.
(b) A contour plot of the density for a stationary obstacle with amplitude $A_p/\mu^*=1.0$.  The shaded region indicates a density exceeding 1.05$n_0$, and the arrow indicates the direction of polarization.
(c) Density slices of (b) along the parallel (dashed red) and perpendicular (blue) directions.  The density oscillation due to the roton is clear in the perpendicular case.
}
\label{fig:bogo}
\end{figure}

To illustrate this point, we use the parameters $g^*=0.25$, $g_d^*=0.36$, $l_z/\xi^*=0.5$ and $\alpha = \pi/4$ ($\mu^*=1$), which are chosen to best illustrate anisotropic effects while keeping safely away from the unstable regime; we will identify them with experimental parameters below.  Figure~\ref{fig:bogo}(a) shows the dispersion calculated using these parameters for quasiparticle propagation parallel to ($\parallel$) and perpendicular to ($\perp$) the tilt of the dipoles into the plane.  For parallel propagation, the dispersion resembles that of a system with contact interactions; the curve goes smoothly from the linear phonon regime at small $k$ to the free-particle regime at large $k$.  For this case, $v_L/c^*=1.0$ (dashed red line), meaning that the critical velocity is identical to the speed of sound.  In contrast, the perpendicular dispersion curve exhibits a roton-like feature at intermediate $k$, setting $v_L/c^*=0.50$ (dashed blue line).  The inset in figure~\ref{fig:bogo}(a) shows $v_L$ as a function of the azimuthal angle, $\eta$, the angle between $\mathbf{k}_d$ and $\mathbf k$.  Interestingly, the speed of sound, given by $c_s=\lim_{k\rightarrow 0}[\omega(k)/k]$, is the same for both parallel and perpendicular propagation, $c_s/c^*=1$, and is in fact isotropic.  Therefore, the anisotropy in the spectrum occurs only at finite $k$ due to the presence of an anisotropic roton. 

The impact of the anisotropic roton can be directly seen in the density of the gas.  In Figure~\ref{fig:bogo}(b) we show a contour plot of the density in the presence of a repulsive Gaussian potential, or ``probe,'' of the form $V_p(x,y)=A_p\exp{(-(x^2+y^2)/\sigma_p^2)}$ with $A_p/\mu^*=1.0$ and $\sigma_p/\xi^* = 2.0$, as may be realized by shining a blue-detuned laser on the system from the $z$-direction.  The  shaded regions indicate density above 1.05$n_0$, and the arrow indicates the direction of polarization.
In Figure~\ref{fig:bogo}(c) we plot density slices of this distribution to more clearly show the density profile in the parallel (dashed red) and perpendicular (solid blue) directions.  Interestingly, the high-density regions occur in the direction perpendicular to the tilt of the dipoles, the same direction that exhibits a roton feature in the dispersion.  Indeed, it was shown in ref.~\cite{Wilson2} that a DBEC in the presence of a repulsive Gaussian (or a vortex core), at sufficiently large density, will exhibit density oscillations due to the manifestation of the roton.  Here, we see a manifestation of the anisotropic roton in the static structure of the q2D DBEC.  

We now address the question of what happens to this anisotropic DBEC when the probe is moved through it with velocity $\mathbf{v}$, by numerically solving Eq.~(\ref{q2DGPE}) with the Gaussian potential $V_p(x-vt,y)$.  For concreteness, we consider motion parallel and perpendicular to the tilt of the dipoles by tilting $\hat{d}$ into the $\hat{x}$ and $\hat{y}$ directions, respectively, while fixing the direction of the probe velocity so that $\mathbf{v}=v\hat{x}$. 

\begin{figure}
\includegraphics[width=0.85\columnwidth]{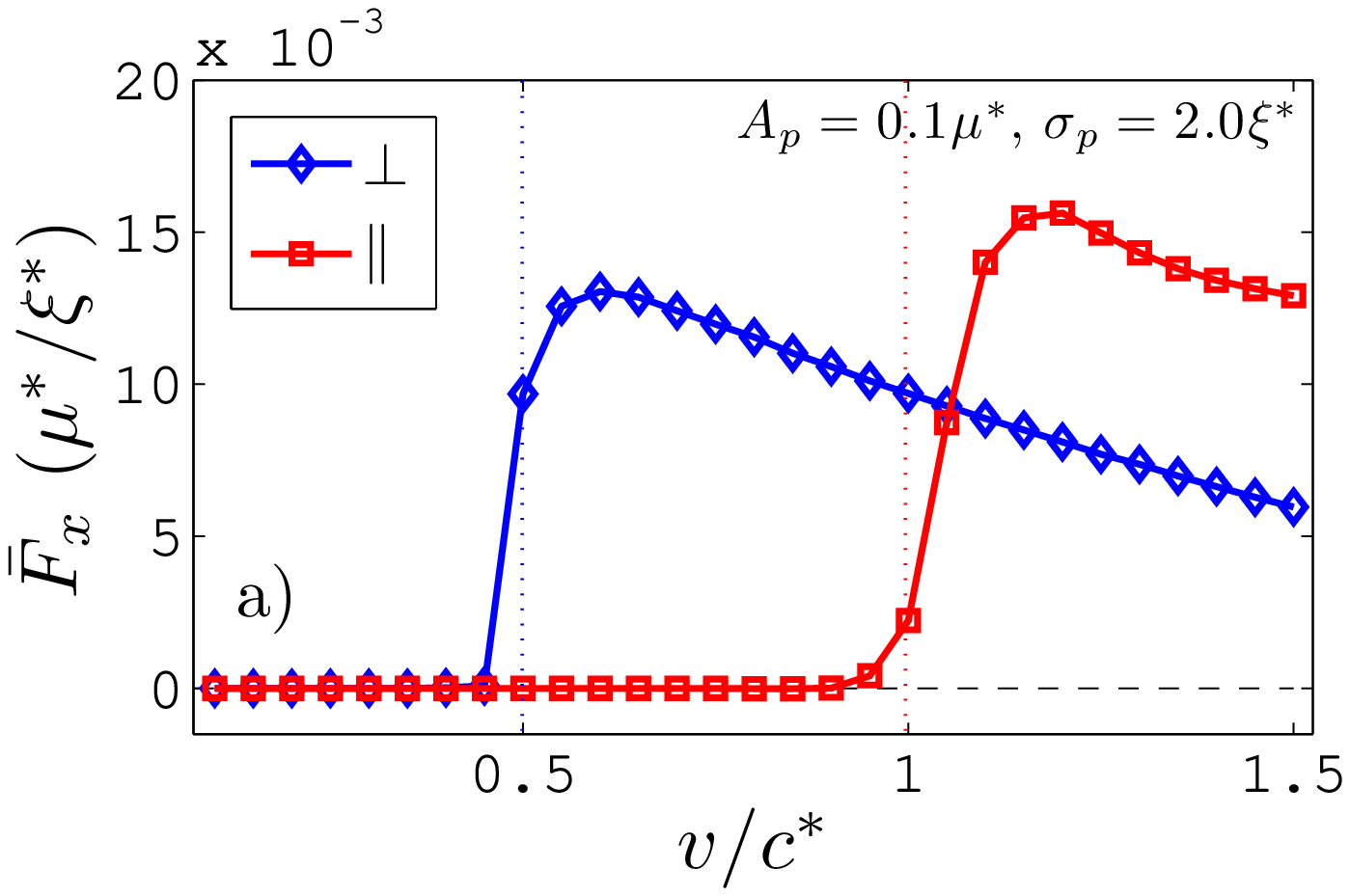} \\
\vspace{-10pt}
\includegraphics[width=0.85\columnwidth]{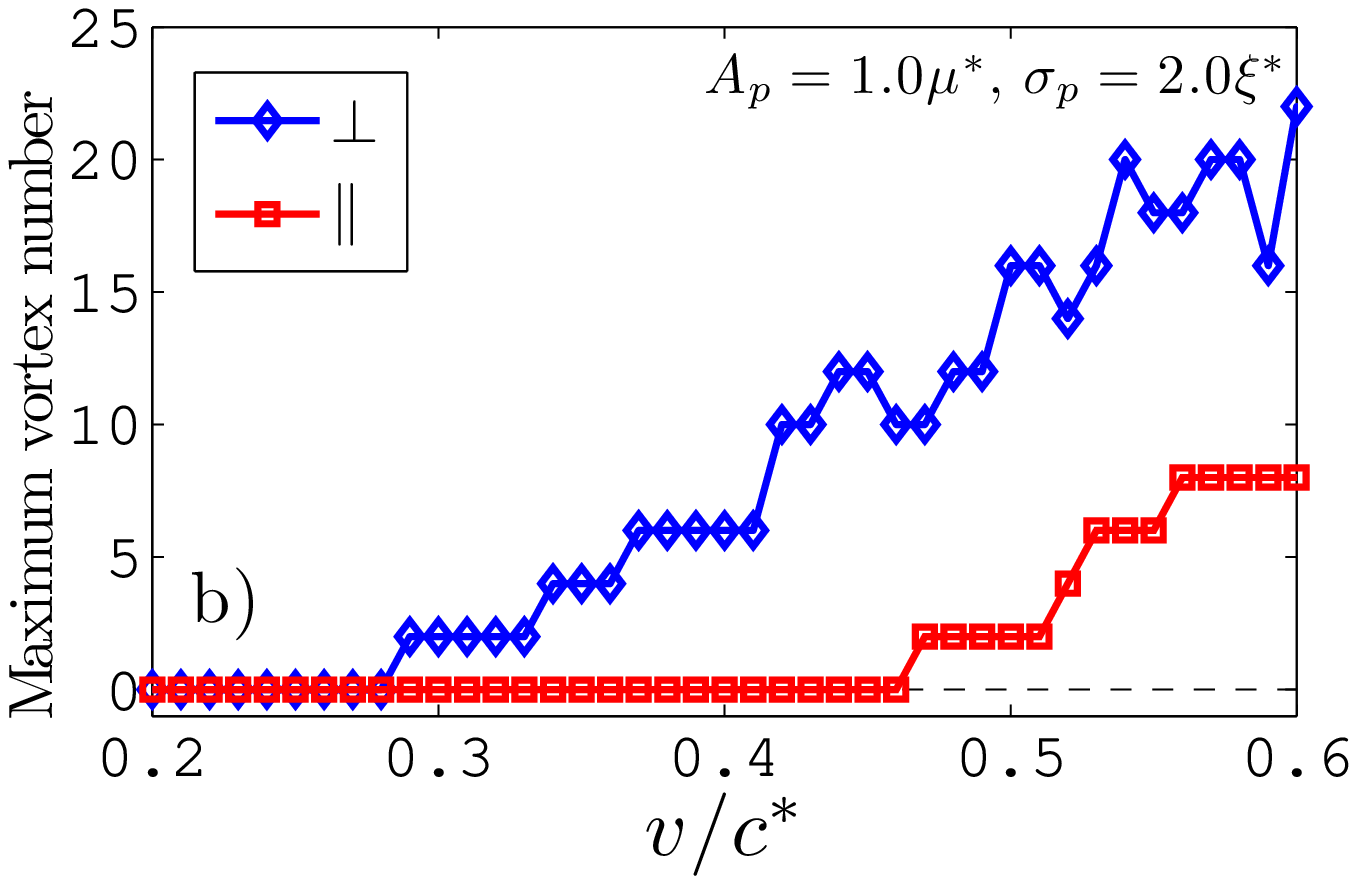}
\vspace{-8pt}
\caption{(color online) 
(a) The mean drag force acting on the probe with $A_p/\mu^*=0.1$ and $\sigma_p/\xi^*=2.0$ calculated up to time $t/\tau^*=100$ for motion perpendicular to (blue diamonds) and parallel to (red squares) the dipole tilt.  The dotted lines represent the corresponding $v_L$, in excellent agreement with the numerical simulations.
(b) The maximum vortex number produced by a probe with $A_p/\mu^*=1.0$ and $\sigma_p/\xi^*=2.0$ calculated up to $t/\tau^*$=100 for motion perpendicular to (blue diamonds) and parallel to (red squares) the dipole tilt.  The corresponding critical velocities are: $v_c^{(\perp)}/c^*=0.27$ and $v_c^{(\parallel)}/c^*=0.46$.}
\label{number}
\end{figure}

Figure~\ref{number}(a) shows the time-averaged drag force (averaged up to $t=100\tau^*$) acting on a ``weak'' probe with parameters $A_p/\mu^*=0.1$ and $\sigma_p/\xi^*=2.0$.  The force at time $t$ is given by $\mathbf{F}(t) = -\int d^2\rho|\psi(\boldsymbol{\rho},t)|^2 \vec{\nabla} V_p(\boldsymbol{\rho},t)$~\cite{win}.  In this case, the probe is sufficiently weak so that no vortices are nucleated in the fluid, and instead only quasiparticles are excited.  The presence of a force on the probe signifies the excitation of quasiparticles, and thus the breakdown of superfluid flow.  There is a clear anisotropic onset of force in these simulations that agrees very well with the anisotropic $v_L$ given by the Bogoliubov dispersions in figure~\ref{fig:bogo}(a), resulting in critical velocities of $v_c/c^*=0.90(0.46)$ for parallel (perpendicular) motion of this probe, determined by the velocity at which the drag force suddenly rises.

It has been shown that $v_L$ is recovered as the true critical velocity only when the superfluid is perturbed by a vanishingly small object~\cite{steissberger,ianeselli}.  Additionally, while quasiparticle excitations are a natural feature to study when considering the breakdown of superfluid flow, they may be difficult to observe experimentally, especially in the limit where the probe is perturbative.  Vortices, on the other hand, are superfluid excitations in the form of topological defects that create regions of zero density and are easier to observe experimentally than quasiparticles. The first measurements of $v_c$ in a BEC were from observations of the sudden onset of heating ~\cite{raman,onofrio}, believed to be related to vortex production in the BECs.  More recently, Ref.\cite{anderson} used experimental finesse to controllably create vortex pairs to observe $v_c$.

Motivated by these circumstances, we investigate the critical velocity for vortex formation in the q2D DBEC by using a moving probe with an amplitude that is linearly ramped from $A_p=0$ to $A_p/\mu^*=1.0$ in a time $10\tau^*$ with $\sigma_p/\xi^* = 2.0$.  The critical velocity in this case corresponds to the probe velocity above which vortices are formed, signaling the breakdown of superfluidity as energy is used to create these topological excitations. 

We observe a significant difference in the critical velocity at which vortices are formed between a probe moving parallel and perpendicular to the dipole polarization.  In Figure~\ref{number}(b) we show the maximum number of vortices formed within $t/\tau^*=100$.  The critical velocities are $v_c=0.46(0.28)$ for motion parallel (perpendicular) to the dipole tilt.  These values are about half the value of the critical velocities obtained using the weaker probe, but this is not unexpected \cite{Frisch,win}.
In a superfluid, vortices have quantized circulation: $\oint \mathbf v \cdot d \mathbf l=2\pi\hbar n/m$, where $\mathbf{v}$ is the velocity field of the fluid and $n$ is an integer, corresponding to phase winding of $2\pi n$ around the vortex core.  We count vortices in our simulations by finding the phase winding on a plaquette of neighboring grid points \cite{foster}.
\begin{figure}
\vspace{3mm}
\includegraphics[width=.9\columnwidth]{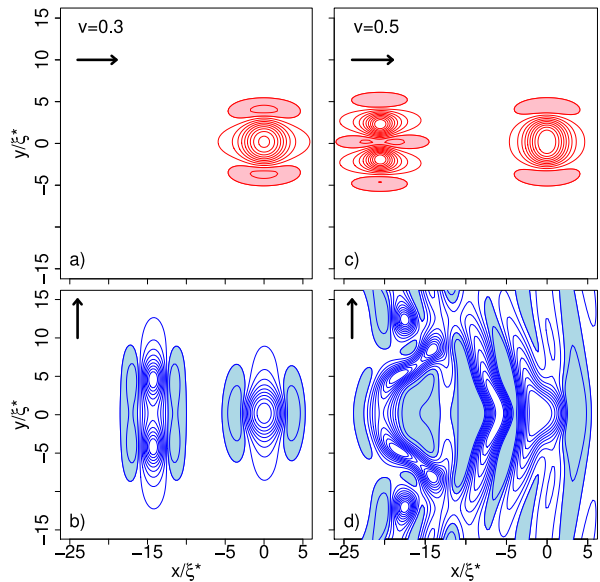}
\caption{(color online) (a) and (b)  The densities are shown for the $\parallel$ (top, red) and $\perp$ (bottom, blue) cases for $v/c^*$=0.30 and $t/\tau^*$=125. Only the $\perp$ case has exceeded its $v_c$. (c) and (d) Densities for $v/c^*$=0.50 at $t/\tau^*$=100.  The
parallel case (top, red) has exceeded its critical velocity and the perpendicular case (bottom, blue) has been wildly excited.  The obstacle is at $x=y=0$ and moving in the $+x$ direction.  The shaded regions of this plot occur when the density exceeds 1.05$n_0$, and the arrow indicates the direction of dipole tilt.}
\label{density}
\end{figure}

The physical mechanism that sets the critical velocity for vortex formation is not rigorously understood.  However, it is theorized that the maximum local fluid velocity about an obstacle, being larger than the background flow velocity, sets the critical velocity via the Landau criterion.  This idea has been fruitful~\cite{win,Frisch}, and we find qualitative agreement with this theory here, as the direction with lower $v_c$ is also the direction of flow most likely to spawn vortices.  However, we note that the ddi is anisotropic although the fully condensed (ground) state of the system is completely isotropic.  The anisotropies only appear in the dispersion relation and in the ground state of the system in the presence of a perturbing potential, which is intimately related to the dispersion relation~\cite{Wilson2}.  Thus, the anisotropies in the critical velocity for vortex formation are due to the anisotropy of the roton mode, just like the critical velocities for quasiparticle excitations.

Figure~\ref{density} shows contour plots of the condensate density for both parallel (red contours, top row) and perpendicular (blue contours, bottom row) motion of the probe relative to the dipole tilt for velocities $v/c^*=0.3$ (left column) and $v/c^*=0.5$ (right column), where the probe is moving in the $\hat{x}$ direction and is located at the origin at the time shown in figure~\ref{density}.  Recall that $v/c^*=0.3$ is just above $v_c$ for vortex formation for perpendicular motion, but well beneath $v_c$ for parallel motion.  This is reflected in the figure, where in (a) no vortices have been formed for parallel motion, while in (b) a vortex pair has been formed for perpendicular motion for the same probe velocity.

For the case of $v/c^*=0.5$, we see that the parallel case in (c) has formed a vortex pair, and in (d) the perpendicular case has been wildly excited.  There is an important contrast to be made in the density profiles when there is a single vortex pair in (b) and (c).  In the parallel case (c) we see that a {\it high} density region occurs between the vortex pair and is elongated in the polarization direction.  In contrast, for the perpendicular case (b) there is a low density region between the vortex pair and high density regions on either side of the vortex pair.  Both the anisotropic superfluid critical velocity for vortex pair production and these contrasting density profiles present means to observe the effects of the roton in DBEC directly.

In addition to investigating this q2D system, we have performed simulations for a fully trapped DBEC.  For a trap aspect ratio of $\lambda = \omega_z/\omega_\rho = 50$, we find critical velocities for vortex production are strongly anisotropic, and the $v_c$ are numerically similar to the free case.  In these simulations, we start the probe in the center of the trap and move outwards in the parallel or perpendicular direction, linearly ramping the amplitude of the laser down to zero by the time it reaches the zero density region.  Such a simulation is experimentally realizable in a DBEC of atomic $^{52}$Cr, for example, having a permenant magnetic dipole moment of $6\mu_B$ where $\mu_B$ is the Bohr magneton, for a DBEC with particle number $N\simeq 18.5\times 10^3$, scattering length $a_s=5.0a_0$ where $a_0$ is the Bohr radius, radial trap frequency ${\omega_\rho}=2\pi\times 20$Hz and a blue-detuned laser with width $\sigma_p=1.76\mu$m.  The speed of sound in this system is $c_s = 0.16$cm/s in the center of the trap~\cite{cr}.

In conclusion, we have characterized the DBEC as an anisotropic superfluid by performing numerical simulations of a blue-detuned laser moving through the system in directions parallel and perpendicular to the dipole polarization.  We find a sudden onset of drag on the laser at velocities that depend strongly on the direction of motion, and attribute the anisotropy in critical velocity to the anisotropic roton so that a measurement of an anisotropic critical velocity in a DBEC corresponds to a measurement of the roton in the system.  Additionally, by considering a DBEC that is experimentally realizable with atomic $^{52}$Cr, we propose a single, stable constituent with which to study anisotropic superfluidity, while other systems such as superfluids of $d$-wave Cooper pairs are more conceptually and experimentally difficult to control.

C.T. gratefully acknowledges support from LANL, which is operated by LANS, LLC for the NNSA of the U.S. DOE under Contract No. DE-AC52-06NA25396. R.M.W. and J.L.B. acknowledge financial support from the DOE and the NSF.

\bibliographystyle{amsplain}

\begin{thebibliography}{99}
\bibitem{Allum} D. R. Allum, P. V. E. McClintock, A. Phillips, and R. M. Bowley, Phil. Trans. R. Soc. A {\bf284}, 179 (1977).
\bibitem{Santos03} L. Santos, G. V. Shlyapnikov and M. Lewenstein, Phys. Rev. Lett., {\bf 90} 250203 (2003).
\bibitem{Wang}D. W. Wang and E. Demler, arXiv:0812.1838v1.
\bibitem{Higbie}J. Higbie and D. M. Stamper-Kurn, Phys. Rev. Lett., {\bf 88}, 090401 (2002).
\bibitem{lattice}I. Danshita and D. Yamamoto, Phys. Rev. A 82, 013645 (2010).
\bibitem{soliton}I. Tikhonenkov, B.A. Malomed, and A. Vardi, Phys. Rev. Lett. {\bf100}, 090406 (2008).
\bibitem{Frisch}T. Frisch, Y. Poeau, and S. Rica, Phys. Re. Lett. {\bf69} 1644 (1992).
\bibitem{win}T. Winiecki, J. F. McCann, and C. S. Adams, Phys. Rev. Lett. {\bf82}, 5186 (1999); T. Winiecki, B. Jackson, J.F. McAnn, and C. S. Adams, J. Phys. B {\bf33}, 4069 (2000).
\bibitem{nath}R. Nath, P. Pedri, and L. Santos, Phys. Rev. Lett. {\bf102}, 050401 (2009).
\bibitem{uwe}U. Fischer, Phys. Rev. A {\bf73}, 031602(R) (2006).
\bibitem{Landau}L. D. Landau, J. Phys. (Moscow) {\bf5}, 71 (1941).
\bibitem{Wilson2}R. M. Wilson, S. Ronen, and J. L. Bohn, Phys. Rev. Lett. {\bf100}, 245302 (2008). 
\bibitem{steissberger}J. S. Stei{\ss}berger and W. Zwerger, Phys, Rev. A, {\bf 62}, 061601(R) (2000).
\bibitem{ianeselli}S. Ianeselli, C. Menotti and A. Smerzi, J. Phys. B., {\bf 39}, S135-S142 (2006).
\bibitem{raman} C. Raman et al., Phys. Rev. Lett. {\bf83}, 2502 (1999).
\bibitem{onofrio} R. Onofrio et al., Phys. Rev. Lett. {\bf85}, 2228 (2000).
\bibitem{anderson}T. W. Neely, E. C. Samson, A. S. Bradley, M. J. Davis, and B. P. Anderson, Phys. Rev. Lett. {\bf104}, 160401 (2010).
\bibitem{foster} C. Foster, P. B. Blakie, and M. J. Davis, Phys. Rev. A {\bf81}, 023623 (2010).
\bibitem{Wilson}R. M. Wilson, S. Ronen, and J. L. Bohn Phys. Rev. Lett. {\bf104}, 094501 (2010).
\bibitem{cr}Th. Lahaye, {\it et al.}, Nature {\bf448}, 672 (2007); Th. Lahaye, {\it et al.}, Phys. Rev. Lett {\bf101}, 080401 (2008).

\end{thebibliography}

\end{document}